\documentstyle[aps,prd,epsf]{revtex}
\begin{document}
\draft
\title{Can dark matter see itself?}

\author{Steen Hannestad}

\address{NORDITA, Blegdamsvej 17, DK-2100 Copenhagen, Denmark}

\date{\today}

\maketitle              

\begin{abstract}
Many independent high-resolution simulations of structure formation
in cold dark matter models show that galactic halos should have singular
core profiles. This is in stark contrast with observations of both
low- and high-surface brightness galaxies, which indicate that the
dark matter has almost constant density in the central parts of halos.
Basically there are three possible avenues to a solution to the problem,
which we discuss in turn. 
Observations of halo profiles could be more uncertain
than previously thought, and higher resolution observations could
reveal that spirals do have a singular core feature. The highest
resolution simulations do not include a baryonic component, and it
is conceivable that violent star formation processes and 
similar phenomena can
destroy the singular dark matter core and lead to an almost constant
density core profile. Finally, we discuss in more detail the intriguing
possibility that the discrepancy hints at some new exotic physics of
the dark matter. Warm dark matter and 
self-interacting dark matter are two of the most promising candidates.

\end{abstract}

\section{Introduction}
Dark matter seems to be a necessary ingredient for structure formation
in the universe \cite{peacock}. 
The standard hierarchical clustering model where
the dark matter is in the form of very non-relativistic, collisionless
particles has been very successful in explaining structure from galactic
scales to the largest scale observable, that of the cosmic microwave
background radiation (CMBR) \cite{peacock,gross}. 
In the 1980s it was realised that the flat, pure CDM
model produces too much small scale structure if it is normalized on
large scales \cite{efst90}. 
Some modification of the model is needed, and several
possible avenues for this exist. 
The most popular model initially was mixed dark
matter \cite{dss92}, 
where some of the dark matter is made up of light particles
with large free streaming length. 
The prime candidate for these light particles is a neutrino with a mass
of a few eV. However, recent results from Super-Kamiokande 
\cite{fuk98} and other
neutrino oscillation experiments suggest that the mass of the most
massive neutrino is of the order 0.1 eV, far too small a mass to be
cosmologically significant. Thus, the mixed dark matter model is currently
disfavoured although perhaps not completely excluded.
However, data from type Ia supernovae at high redshift suggests that
the energy density in the universe is dominated by a cosmological
constant \cite{riess98,perl99}. 
This also dampens small scale structure formation and can
bring the CDM model into agreement with observations without the need
for hot dark matter.

Recently, however, a new problem for the CDM model has surfaced,
having to do with structure on galactic scales and below.
Observations of spiral galaxies, both high and low surface brightness,
suggest that the dark matter in spirals is in the form of a halo with almost
constant central density \cite{bm97,mb98,pick97,sb00,bsal00}. 
To a reasonable approximation it can
be described by a universal profile of the form \cite{bsal00}
\begin{equation}
\rho(r) \simeq \rho_0 \frac{r_0^3}{(r+r_0)(r^2+r_0^2)}
\to_{r\to 0} \rho_0,
\end{equation}
where the central density, $\rho_0$, and the core radius, $r_0$,
depend on the specific system.
However, numerical simulations of halo formation consistently show
much more singular core profiles. The first large scale study of
CDM halo profiles, with sufficient resolution to resolve 
galaxy sized halos, was that of Navarro, Frenk and White
\cite{NFW96}, who found that
the simulated CDM halos also follow a universal profile
\begin{equation}
\rho(r) \simeq \frac{\rho_c \delta_c}{(r/r_s)(1+r/r_s)^2}
\to_{r\to 0} \rho_c \delta_c \frac{r_s}{r} ,
\end{equation}
known as the NFW profile. 
$\rho_c \delta_c$ and $r_s$ are again parameters that are to
be fitted to individual systems.
Close to the center this profile approaches
$\rho \propto r^{-1}$, in strong disagreement with observations.
More recent simulations with much higher resolution find an even
steeper profile, approaching $\rho \propto r^{-3/2}$
\cite{FP94,N99,NS99,moore2}. 
Furthermore, it has been claimed that these CDM simulations are 
now of sufficiently high resolution that the results for the central halo 
profile have converged to the infinite resolution limit.

In addition to this problem with the halo density profiles,
observations show that a galaxy like the
Milky Way contains almost an order of magnitude fewer small satellite
galaxies than what is found in CDM simulations \cite{moore,ghigna}.

\section{Possible solutions}

This discrepancy between CDM simulations and observations of galaxies
is so serious that it needs to be investigated more closely, because
it could indicate a fundamental flaw in our understanding of structure
formation and the nature of dark matter. 
There are basically three possible ways to remedy
the discrepancy, which we shall discuss in turn.

\subsection{Observational solutions}

Until very recently, relatively little effort has been devoted to measuring
reliable rotation curves in the inner parts of low surface brightness
(LSB) galaxies,
the reason being that the outer parts of the rotation curves are 
completely dominated by dark matter, and thus hold important information
about the total dark matter mass in the halo. Even if there is a 
singular core in the halo, its integrated mass is only a small fraction
of the total halo mass. In that case, high angular resolution in
the central part of the galaxy is not an important issue.
Therefore
observations of galactic rotation curves have primarily been made
using observations of the 21cm neutral hydrogen hyperfine-structure
line with radio telescopes.
These observations have the advantage that neutral hydrogen
systems extend further out in most galaxies than the visible
light, meaning that the halo structure can be mapped out to
very large radii. If the total halo mass is the parameter
to be determined then this technique is indeed the best possible.
But because of the poor angular resolution of radio observations there
are indeed significant effects of beam-smearing in the measurements
of the inner rotation curves.
Furthermore there is usually not
much neutral hydrogen present within the central kpc of a galaxy.

Recently it was pointed out by van den Bosch and Swaters \cite{bs00} that
taking beam smearing into account increases the uncertainty in
rotation curve measurements by a significant amount. In fact it seems
that some LSB galaxies are consistent with having a cusp-like feature
in the central halo, whereas others are still best fitted by a constant
density core. At present it is not clear how important this
beam-smearing effect really
is, and other very recent studies reach the opposite conclusion, that
galaxy halos are best fitted by constant density cores \cite{sb00}.

Instead of using radio observations to probe the inner parts of
rotation curves
it may be more feasible to use observations of HII regions
instead of neutral hydrogen. These observations can be made with
optical telescopes and in principle have very high angular
resolution. HII is usually also very abundant in the central parts
of galaxies. One very recent study of HII rotation curves 
by Swaters, Madore and Trewhella \cite{smt00} indicates
that there could be cusps present, but on the other hand they also
do not rule out the constant density cores.
Clearly, better observational data from the inner parts of galaxies are
needed, as well as an increased understanding of the possible 
uncertainties involved in these observations.

It is more difficult to imagine an observational solution to the
substructure problem. If small satellite galaxies were present in galactic
halos in the numbers predicted by CDM simulations they would doubtless
have been detected. However, that conclusion assumes that their mass-to-light
ratio is not too different from ordinary galaxies.
As will be discussed in the next section, it has been proposed that small
satellite galaxies are not seen, simply because they have extremely
large mass-to-light ratios.

\subsection{Astrophysical Solutions}

CDM simulations with very high resolution (sufficient to resolve 
cores of galactic halos) at present do not include baryons
\cite{moore,moore2}.
The reason is both that the computational demands for this type of 
simulation are very high, and that there is no really reliable way
of including star formation, supernovae and similar small scale
features in the baryonic component.
Without such high resolution simulations it is difficult to completely
exclude the possibility that a proper treatment of the baryons would
cure the discrepancy between simulations and observations.

Several scenarios have in fact been proposed that try to remedy the 
problem by invoking baryonic features.
As was mentioned in the last section, small satellite galaxies would
not have been detected if they contain very few stars.
Bullock, Kravtsov and Weinberg \cite{bkw00}
have suggested that when the universe
was reionised at a redshift $z \gtrsim 5$, accretion of baryons onto
existing low mass dark matter halos was almost completely halted.
The presence of a strong ionising radiation field stops formation
of galaxies with velocity dispersion smaller than $\simeq 30$ km/s,
almost exactly the scale below which the discrepancy between the
predicted and observed number of systems becomes apparent.
If this explanation is true, then the prediction is that galactic
halos contain a very large number of dark matter clumps with very few stars
in them. Such clumps could possibly have disastrous effect on disk
galaxies. It is well known that thin disks are unstable to heating by
the frequent passage through the disk of massive objects, whereas 
thick disks are more robust. At present it is not clear how the
distribution of dark matter clumps predicted by CDM would affect 
disk galaxies, and better simulations of this effect would be highly
desirable.

It has also been proposed that tidal interactions between baryons and
dark matter in the central halo can erase the cusp-like feature seen
in pure CDM simulations \cite{bgs00}. 
In this scenario a singular CDM halo forms
at high redshift and starts accreting a baryonic disk. At some point
a massive burst of star formation and subsequent supernova events
occurs, initiating a galactic wind that can eventually carry off
almost all gas present in the disk. If not too much star formation
has occurred prior to the wind formation, the total baryonic mass
carried off could be close to the total mass of the disk.
This dramatically changes the gravitational potential felt by the
dark matter and the subsequent violent relaxation can erase the singular
core. After this event, the baryonic disk which is observed in present
day spirals is slowly accreted at lower redshift.

Even though this scenario seems plausible there are a number of serious
questions related to it. Firstly, preliminary results of galactic winds
driven by star formation show that almost all metals can be easily
expelled from the galaxy because of their high opacity \cite{mf99,eb99}. 
This fits well with
the fact that metals are observed in abundance in the intergalactic
medium at high redshift. However, it seems impossible to expel a large
fraction of the hydrogen, which makes up most of the mass
\cite{mf99,eb99}. Thus it
seems unlikely that the required amount of baryons can be expelled from
the galaxy by this initial burst of star formation.
Secondly, the hypothesis that most of the baryons remain in diffuse
gas until they are expelled is neither supported nor ruled
out by present observations. In order to determine if this is indeed the
case, a better understanding of star formation is needed.

Thus, the final conclusion is that a scenario where the problem is resolved
by properly taking into account interactions between baryons and the
dark matter cannot be ruled out at present. However, explaining the
cusp problem requires some fairly strong assumptions about the behaviour
of the baryonic component at high redshift.

\subsection{Possible particle physics solutions}
If the astrophysical or observational solutions turn out not to work, then
the only option we have left is to consider solutions from particle
physics. Roughly this type of solution can be divided up into two categories,
those that derive from physics at the epoch of initial power spectrum 
formation and those which come from changing the physical properties of 
the dark matter particles.

The first type of solution is in a sense the simplest because it does
not affect physics during structure formation, only the initial
conditions. The initial condition in this regard is the power spectrum
of fluctuations from inflation. In order to solve the substructure
problem we need to dramatically reduce fluctuations on scales smaller
than a galaxy, or
\begin{equation}
\frac{k}{1 \,\, {\rm Mpc}^{-1}} \gtrsim 5 \, \left(\frac{M}{10^{11}M_{\odot}}
\right)^{1/3}.
\end{equation}
This could for instance be the case in inflationary models with broken
scale invariance, as was proposed by Kamionkowski and Liddle \cite{KL99}.
These authors showed that the substructure problem could indeed be solved
by this method. However, the cusp problem is more serious. It has been shown 
in N-body CDM simulations, with an artificial power spectrum cut-off 
introduced,
that dark matter halos also have singular cores \cite{moore2}. 
The reason for this is
simply that in any cold, collisionless collapse there is a large amount
of material with very low entropy (temperature). This material eventually
ends up in the centers of gravitational wells, forming the singular cores 
observed in simulations. Thus, the conclusion is that introducing a power
spectrum cut-off does not remedy the problem, it is a generic feature
in any collisionless CDM structure formation scenario, regardless of the
initial conditions.

This leads us to consider the other possible solution which has to do with
the nature of the dark matter itself.
A very interesting possibility was proposed by Spergel and Steinhardt
\cite{ss99} (see also \cite{han99,bur00,fir00,yos00,mor00,ost99,mir00}),
namely that the cold dark matter could possess relatively strong 
self-interactions. If there are such self-interactions they have significant
effect on halo formation. For example substructure clumps will be 
evaporated by interactions with the smooth dark matter halo background,
and could decrease the number of substructure halos to the observed
level.

The maximum entropy state of any self-gravitating system is the singular
isothermal sphere, regardless of possible self-interactions. However,
if the interaction cross-section has the right magnitude, collapse to
a singular halo could be prevented for a sufficiently long time that
present day halos would not show singular cores. It has been shown in
numerical simulations \cite{wan00,dav00,yos200} 
that a cross section higher than 
\begin{equation}
\frac{\sigma}{m} \gtrsim 10^{-23}-10^{-22} \,\, {\rm cm}^2/{\rm GeV}
\end{equation}
leads to quick core-collapse and formation of a singular core. On the
other hand, cross sections smaller than 
\begin{equation}
\frac{\sigma}{m} \lesssim 10^{-25} \,\, {\rm cm}^2/{\rm GeV}
\end{equation}
will not have an observable effect on structure formation. If the cross 
section is fine-tuned to a value
\begin{equation}
\frac{\sigma}{m} \gtrsim 10^{-24}-10^{-23} \,\, {\rm cm}^2/{\rm GeV},
\end{equation}
the initial expansion of the halo core prior to collapse lasts more than
a Hubble time, so that present day halos could fit observations
(see A.~Burkert in this volume for a more thorough discussion of this
model). At present it is not entirely clear whether the scattering
cross-sections needed to explain the substructure problem are the 
the same as are needed to explain the cusp problem. If this turns out
not to be the case, then self-interacting cold dark matter is certainly
disfavoured.

Another possibility which has received a lot of attention recently
is that dark matter is not completely cold, but has some thermal motion
in the early universe \cite{HD,hd00,SD99,NSD,CAV}. 
If the dark matter has thermal motion, any
fluctuations are erased on scales smaller than the free-streaming length
\cite{peacock}
\begin{equation}
k_{\rm streaming} \simeq 1 \,\, {\rm Mpc}^{-1} \, (m/1 \, {\rm keV})^{3/4},
\end{equation}
for particles which are in a relativistically decoupled thermal 
distribution.
In the old hot dark matter model of structure formation, where the dark
matter was in the form of neutrinos with a mass of $\sim 30$ eV, the
free-streaming length would be of order $k_{\rm streaming} \simeq 0.05 
\,\, {\rm Mpc}^{-1}$, roughly the size of a large cluster.
This is clearly too large to fit large scale observations, but if the
mass of the dark matter particle is $\sim 1$ keV, the free-streaming length
is of the order $k_{\rm streaming} \simeq 1-2 \,\, {\rm Mpc}^{-1}$,
just what is needed to explain the substructure problem.
At first sight this ``warm'' dark matter (WDM) model 
\cite{HD,hd00,SD99,NSD,CAV,SS88,CSW96,burns}
would seem to suffer from 
the same problem as those with a power spectrum cut-off, namely that
low entropy material clusters in the central parts of halos and produce
singular cores. However, there is one vital difference between this
model and CDM with a power spectrum cut-off. The warm dark matter
has finite temperature, which implies that the phase-space distribution
occupies a finite volume in momentum space. On the other hand,
the CDM distribution is by definition a $\delta$-function in momentum
space.
A useful quantity describing this is the average phase space density
\cite{HD,hd00}
\begin{equation}
Q \equiv \frac{\rho}{\langle v^2 \rangle^{3/2}}
\cases{< \infty & for WDM \cr \equiv \infty & for CDM}.
\end{equation}
In an adiabatic collapse to a lowered isothermal sphere, the
central halo density can be crudely estimated as \cite{HD,hd00}
\begin{equation}
\rho_0 \simeq Q (3 \sigma)^{3/2}
\cases{< \infty & for WDM \cr \equiv \infty & for CDM}.
\label{eq:core}
\end{equation}
Thus, WDM halos are expected to have non-singular cores. However, the 
question is whether the same WDM particle mass can solve both the substructure
problem and the cusp problem.
From studies of galactic substructure, as well as the Lyman-$\alpha$
forest \cite{NSD} 
it has been concluded that $m_{\rm WDM} \gtrsim 750 \,\, {\rm eV}$ in
order to fit observations.

To date no N-body simulation with sufficient resolution to tackle
the WDM effect on halo cores has been performed. We are therefore for the
moment forced to rely on crude estimates of the effect, such as 
Eq.~(\ref{eq:core}). Hogan and Dalcanton \cite{HD,hd00}
have estimated that the
WDM mass should be $m_{\rm WDM} \simeq  200-300 \,\, {\rm eV}$ in order to
produce the correct core radii for low surface brightness galaxies.

It therefore seems that the collisionless WDM model has difficulty
in explaining both the substructure and the cusp problem
(however, see also Ref.\ \cite{madsen}). 
However, it
should not be immediately dismissed because whereas the estimate of the
mass needed to reproduce substructure is probably quite accurate,
the estimate of the mass needed to resolve the cusp problem is at best
a crude guess.

If it turns out that collisionless WDM does not work then WDM with
self-interactions might \cite{hs00}. 
Even relatively weak self-interactions will
keep the WDM distribution in pressure equilibrium until $T \ll m$ in
the early universe. In that case the WDM particles cannot free-stream
and perturbation on small scales are not damped. However, the model does not
resemble CDM either because the pressure equilibrium prevent fluctuations
smaller than the Jeans mass from growing \cite{CMH92,machacek,LSS}. 
Since the Jeans length is almost a factor of 2 smaller than the free 
streaming scale \cite{hs00}
\begin{eqnarray}
k_J & \simeq & 1.7 \,\, m_{\rm keV}^{3/4} \,\,\, {\rm Mpc}^{-1} \\
k_{\rm stream} & \simeq & 1.1 \,\, m_{\rm keV}^{3/4} \,\,\, {\rm Mpc}^{-1},
\end{eqnarray}
we find that the mass needed to explain the substructure problem in
this model is only $m \simeq 300-400$ eV, which is quite close to the
mass needed to explain finite cores.

\section{Discussion}

We have reviewed the nature of the apparent discrepancy between CDM
simulations of galactic halos and observations of spiral galaxies.
In addition we have discussed which possible solutions there are
to this problem. The simplest possibility is that either observations
or simulations are in error, but presently it is not at all clear whether
or not this is indeed the case. If it turns out that the discrepancy 
persists even with better simulations and observations, the only other
solution is some additional particle physics. Merely changing the
initial power spectrum from inflation does not work, so the only option
is that the dark matter is not standard CDM. 
Both self-interacting CDM and warm dark matter are intriguing 
possibilities, although at the moment these models have not been sufficiently
investigated to determine whether they work better than CDM.
In any case, the CDM problem on small scales has stirred a lot of 
interest recently and is likely to do so in the future as well.

\end{document}